\documentclass{nature}
\usepackage{color}
\usepackage{paralist}
\usepackage{amssymb,amsmath}
\usepackage{notes2bib}
\title{Direct observation of ultrafast long-range charge \\ separation at polymer:fullerene heterojunctions} 
\author{Fran\c{c}oise~Provencher,$^{\ref{UdeM}}$ Nicolas~B\'erub\'e,$^{\ref{UdeM}}$ Anthony~W.~Parker,$^{\ref{RAL}}$ Gregory~M.~Greetham,$^{\ref{RAL}}$ Michael~Towrie,$^{\ref{RAL}}$ Christoph~Hellmann,$^{\ref{ICLmat}}$ Michel~C\^ot\'e,$^{\ref{UdeM}}$ Natalie~Stingelin,$^{\ref{ICLmat}}$ Carlos~Silva$^{{\ref{UdeM}},\ref{ICLphys}}$ and Sophia~C.~Hayes$^{\ref{UCy}}$}

\begin{document}

\maketitle

\begin{affiliations}
\item D\'epartement de physique \& Regroupement qu\'eb\'ecois sur les mat\'eriaux de pointe, Universit\'e de Montr\'eal, C.P. 6128, Succursale centre-ville, Montr\'eal (Qu\'ebec), H3C~3J7, Canada\label{UdeM}
\item Central Laser Facility, Research Complex at Harwell, Science and Technology Facilities Council, Rutherford Appleton Laboratory, Harwell Oxford, Didcot, Oxfordshire OX11~0QX, United~Kingdom\label{RAL} 
\item Department of Materials and Centre for Plastic Electronics, Imperial College London, South Kensington Campus, London SW7~2AZ, United~Kingdom\label{ICLmat}
\item Visiting Professor (Experimental Solid State Physics), Department of Physics, Imperial College London, South Kensington Campus, London SW7~2AZ, United~Kingdom\label{ICLphys}
\item Department of Chemistry, University of Cyprus, P.O. Box 20537, 1678 Nicosia, Cyprus \label{UCy}
\end{affiliations}

\newpage 

%\section*{Abstract} %typically 150 words
\begin{abstract}
In polymeric semiconductors, charge carriers are polarons, which means that the excess charge deforms the molecular structure of the polymer chain that hosts it. This effect results in distinctive signatures in the vibrational modes of the polymer. We probe polaron photogeneration dynamics at polymer:fullerene heterojunctions by monitoring its time-resolved resonance-Raman spectrum following ultrafast photoexcitation. We conclude that polarons emerge within 200\,fs, which is nearly two orders of magnitude faster than exciton localisation in the neat polymer film. Surprisingly, further vibrational evolution on  $\mathbf{\lesssim 50}$-ps timescales is modest, indicating that the polymer conformation hosting nascent polarons is not significantly different from that in equilibrium. This suggests that charges are free from their mutual Coulomb potential, under which vibrational dynamics would report charge-pair relaxation. Our work addresses current debates on the photocarrier generation mechanism at organic semiconductor heterojunctions, and is, to our knowledge, the first direct probe of molecular conformation dynamics during this fundamentally important process in these materials.
\end{abstract}

\newpage

%S : Excitonic solar cells separate charges
In photovoltaic diodes based on blends of polymeric semiconductors and fullerene derivatives, photocurrent generation requires charge separation, with photoinduced electron transfer as an early step. From a semiconductor perspective, the electronic structure of the two materials define a type-II heterojunction, providing sufficient driving force to dissociate highly-bound excitons on the polymer.\cite{Clarke:2010gb} In spite of important progress to optimise the solar power conversion efficiency in polymer solar cells, the mechanism for the evolution from the non-equilibrium primary photoexcitation (the unrelaxed exciton on a polymer chain immediately after light absorption) to photocarriers (unbound charges) is not resolved and is currently the subject of vivid debate.\cite{Provencher:fk} Earlier studies of all-polymer donor:acceptor blends by Morteani~\textit{et~al}.\ concluded that an intermediate step between intrachain excitons and free charges is charge separation into Coulomb-bound electron-hole pairs,\cite{MorteaniPRL2004} which we will refer to as charge-tranfer states (CT), emphasising that the electron and hole are mutually bound across the interface. After this initial step, the CT state can either branch directly to free charges or relax to an excitonic state bound at the interface between the donor and acceptor materials, a state termed charge-transfer excitons (CTX). Others also reported this branching behaviour in polymer:fullerene blends.\cite{Hallermann:2008dj,Bakulin2009} Later reports indicated that initial charge separation to CT states occurs $\leq 100$-fs timescales.\cite{Tong2010, Grancini2012} Furthermore, Jailaubekov~\textit{et~al}.\ concluded that at molecular-semiconductor heterojunctions, CT states must produce photocarriers on timescales faster than $\sim 1$\,ps if this process is to be competitive against relaxation to the CTX state,\cite{Jailaubekov2012} as the binding energy of CTX is typically $\sim 10\,k_B T$ at room temperature,\cite{Hallermann:2009cq,Gelinas:2011cp} rendering photocarrier generation from those states energetically unlikely in principle. Nevertheless, these relaxed CTX can be `pushed' with an infrared optical pulse resonant with an intraband polaronic optical transition to promote them to a delocalised, near-conduction-edge state that can then dissociate into charge carriers, thus enhancing photocurrent.\cite{Bakulin2012} However, Lee~\textit{et~al}.\cite{Lee2010} and more recently Vandewal~\textit{et~al}.\cite{Vandewal:2013uq} have argued that relaxed CTX states, when excited directly, are in fact the photocurrent precursors in many if not most polymer:fullerene systems. The contemporary literature therefore reflects disagreement on whether rapid dissociation of non-equilibrium intrachain excitons or charge-transfer excitons is a significant pathway towards photocurrent generation in polymer diodes.
    
The seemingly conflicting scenarios are all based on probing population dynamics by time-resolved spectroscopies, or directly probing photocurrent internal quantum efficiency spectra in conjunction with steady-state absorption and luminescence measurements, both of which provide limited direct mechanistic insight into photocarrier generation dynamics. In this work, we exploit an ultrafast optical probe that is simultaneously sensitive to molecular-structure and electronic-population (excitons, polarons) dynamics. It is therefore an intricate probe of charge photogeneration dynamics since charges in polymeric semiconductors are polaronic, in which the charge distorts the polymer backbone surrounding it, producing unambiguous molecular vibrational signatures. 
Although ultrafast solvatochromism assisted vibrational spectroscopy has been used to study charge dynamics in fullerene aggregates,\cite{Pensack2009JACS} our work is to our knowledge the first direct time-resolved vibrational probe of polaron formation dynamics on polymers at polymer:fullerene heterojunctions.  
%\bibnote{We note that charge dynamics in organic-semiconductor films have been probed by ultrafast solvatochromism assisted vibrational spectroscopy in fullerene aggregates; Pensack, R.\ D.\ \& Asbury, J.\ B.\ Beyond the Adiabatic Limit: Charge Photogeneration in Organic Photovoltaic Materials. \textit{J. Phys. Chem. Lett.} \textbf{1}, 2255--2263 (2010).}
We find that by optical excitation of the lowest ($\pi, \pi^{*}$) optical transition in the polymer, clear polaronic signatures are present within the early-time evolution of the photoexcitation, on $\lesssim 200$-fs timescales, in a benchmark polymer:fullerene blend. Surprisingly, we observe limited vibrational relaxation following this ultrafast process,  indicating little structural changes occuring on time windows spanning up to 50\,ps, which we interpret as indicative of holes on the polymer that are sufficiently far from their electron counterparts such that they are free of their Coulomb potential. Our results suggest that in this system, which produces amongst the highest efficiency solar cells, intermediate charge-transfer states (Coulomb-bound electron-hole) pairs are not prerequisite species for photocarriers. Morevoer,  because vibrational signatures of polarons emerge at a faster rate than exciton relaxation by at least two orders of magnitude, we conclude that ultrafast dissociation of non-equilibrium photoexcitations is a significant process in this system. %Furthermore, the modest subsequent conformational reorganisation implies that the ground-state conformation is that required to host the photogenerated hole.

We implement femtosecond stimulated Raman spectroscopy (FSRS) to characterise the molecular structural dynamics in a model semiconductor copolymer used in efficient solar cells during charge transfer to a fullerene acceptor in the solid state.\cite{McCamant2004} FSRS measures the transient resonance stimulated Raman spectrum of the photoexcited species by preparing the polymer in its excited state with a short actinic pulse ($\sim 50$\,fs, 560\,nm), and then probing the evolution of the stimulated Raman spectrum with a pair of pulses: a Raman pump pulse resonant with a photoinduced absorption of interest (1.5\,ps, 900\,nm) and a short ($\sim 50$\,fs) broadband probe pulse to generate the excited-state stimulated Raman spectrum of the transient species at various delay times between the actinic and Raman-pump/probe pulses. FSRS thus permits measurement of the time-resolved, excited-state resonance-Raman spectrum with ultrafast time resolution ($\sim 70$\,fs). We collect concomitantly transient absorption spectra by monitoring the differential transmission of the broadband probe pulse induced by the actinic pulse (i.e.\ by measuring the signal without the Raman pump pulse). In our studies, we apply this technique on poly(N-9$^\prime$-heptadecanyl-2,7-carbazole-alt-5,5-(4$^{\prime}$,7$^\prime$-di-2-thienyl-2$^\prime$,1$^\prime$,3$^\prime$-benzothiadiazole) (PCDTBT, see Fig.~\ref{fig:fig1}a), which is a carefully engineered `push-pull' material system for solar cells, with reported solar power conversion efficiency of 6\cite{Park2009} -- 7.2\%,\cite{Sun2011} unusually high internal quantum efficiency approaching 100\%,\cite{Park2009} and reported ultrafast charge separation in fullerene blends,\cite{Tong2010,Etzold2011} such as fullerene derivative
[6,6]-phenyl-C$_{61}$ butyric acid methyl ester (PCBM, see Fig.~\ref{fig:fig1}a).

%%%%%%%%%%%%%%%%%%%%%%%%%%%%%%%%%%%%%%%%%%%%%%%%%%%%%
\begin{figure}
\caption{\textbf{$| \;$ Steady-state absorption, transient absorption, and Raman spectra of films of neat PCDTBT and PCDTBT:PCBM.} (\textbf{a}) Structure of PCDTBT (left) and PCBM (right). (\textbf{b}) Absorption spectra of neat PCDTBT (blue), PCDTBT:PCBM (pink) and doped PCDTBT (grey) films. (\textbf{c}) Transient absorption spectra of the same samples as in part (a). (\textbf{d}) Resonance Raman spectra of the same samples as in part (a). The wavelength of the actinic (the femtosecond pump) and Raman (picosecond pump) pulses are indicated by vertical lines in part (b).
\label{fig:fig1}}
\end{figure}
%%%%%%%%%%%%%%%%%%%%%%%%%%%%%%%%%%%%%%%%%%%%%%%%%%%%%%

Unequivocal identification of polarons in PCDTBT requires knowledge of the spectral signatures of its cations. To measure them, we oxidised a neat polymer film, that is a film composed only by the polymer material, with FeCl$_3$, which is known to dope polycarbazoles.\cite{Aich2009} The ground-state absorption spectrum of the doped film (Fig.~\ref{fig:fig1}b) features a low-energy absorption band centred around 1.7\,eV, which has been assigned to cations in previous doping studies.\cite{Aich2009} This absorption band in the doped film corresponds to the `interband' polaronic transition.\cite{Wiebeler2013} It matches closely the transient absorption band of the PCDTBT:PCBM blend in the 1.2--2.0-eV region (Fig.~\ref{fig:fig1}c), which reinforces previous assignment of this photoinduced absorption band to polarons.\cite{Tong2010, Etzold2011} Excitation of the doped PCDTBT film within the low-energy band at 830\,nm (1.49\,eV) provides thus the resonance-Raman spectrum of the polaron (Fig.~\ref{fig:fig1}d, grey), which is the vibrational fingerprint of that photoexcitation on PCDTBT. Fig.~\ref{fig:fig1}d also shows the resonance-Raman spectra of neat PCDTBT and PCDTBT:PCBM, excited at 514\,nm (2.30\,eV). To aid with assignment of these peaks, we have calculated Raman frequencies of the ground, excited and cationic states of the PCDTBT dimer in vacuum using density-functional theory (DFT) and time-dependent density-functional theory (TDDFT) utilizing the B3LYP functional. We then identified the modes of the measured Raman peaks by comparing them to the frequency and intensity of the bands in the theoretical Raman spectrum (see Fig.~S2 in Supporting Information) and to a published Raman study of PCDTBT.\cite{Reish2012} These assignments are summarized in Tables 1 and 2.

The difference in intensities of the cation resonance-Raman spectrum with respect to that of the pristine (as-cast) film, consisting of neutral polymer chains, reveal differences in the geometric distortion between the ground and the resonant-electronic states along the normal coordinate for each mode. In particular, the Raman spectrum on resonance with the polaronic transition shows a marked enhancement of bands corresponding to carbazole and thiophene local modes between 1000 and 1200\,cm$^{-1}$, as well as in benzo\-thia\-diazole and carbazole bands at 1268 and 1349\,cm$^{-1}$, which are more  delocalised (see Fig.~S3 in Supporting Information). This is consistent with the electronic density shift between the highest occupied molecular orbital (HOMO) and the lowest unoccupied molecular orbital (LUMO) of the charged polymer, which affects mainly the carbazole and the thiophene units (see Fig.~S10 in Supporting Information). On the other hand, the resonance-Raman spectra of PCDTBT and PCDTBT:PCBM exhibit strong enhancement of modes predominantly involving motion of the benzothiadiazole unit at 1370\,cm$^{-1}$ and 1540\,cm$^{-1}$. This is consistent with the electronic density shift between the HOMO and the LUMO of the neutral polymer, which involves the localisation of the wavefunction on the thiophene--benzothiadiazole--thiophene unit (see Fig.~S13 in Supporting Information).

%%%%%%%%%%%%%%%%%%%%%%%%%%%%%%%%%%%%%%%%%%%%%%%%%%%%%%%
\begin{figure}
\caption{\textbf{$| \;$ Excited-state transient resonance-Raman spectra of PCDTBT:PCBM and neat PCDTBT.} Shown are femtosecond stimulated Raman spectra of (\textbf{a}, \textbf{b}) PCDTBT:PCBM and (\textbf{c}) neat PCDTBT films to illustrate structural relaxation over a 100-ps time window. Spontaneous resonance Raman of doped PCDTBT (green) is also shown in parts (a) for comparison.  
\label{fig:fig2}}
\end{figure}
%%%%%%%%%%%%%%%%%%%%%%%%%%%%%%%%%%%%%%%%%%%%%%%%%%%%%%%
          
We now investigate whether the positive polaron vibrational signature appears in the FSRS spectrum of a PCDTBT:PCBM film, whose evolution is illustrated in Figs.~\ref{fig:fig2}a and \ref{fig:fig2}b (see Fig.~S8 and S9 in Supporting Information for FSRS spectra at all recorded time delays).
We use a Raman pump pulse at 900\,nm (1.38\,eV) to be on resonance with the red edge of the polaron optical absorption, and thus these figures represent the temporal evolution of the stimulated resonance-Raman spectrum of the photoinduced polarons in PCDTBT:PCBM heterojunctions. We note the striking similarity of the transient FSRS spectra with the steady-state, spontaneous resonance-Raman spectrum of the doped PCDTBT film, confirming that we are probing hole polarons directly on the PCDTBT backbone. The characteristic bands of this polaronic signature, in particular the 1364\,cm$^{-1}$ band that is absent from the FSRS exciton spectrum (see Fig.~\ref{fig:fig2}c),  readily follow the ultrashort actinic pulse (see also Fig.~S1 in Supporting Information). 
The clear observation of the cation Raman spectrum at very early times is possible because vibrational coherence dephasing is fast, thus minimising parasitic `derivative' line shapes. Indeed, FSRS is a coherent spectroscopy that is able to monitor changes in vibrational frequency that occur faster than the vibrational dephasing time.  This vibrational dephasing is portrayed within the spectra by phase shifts that transpose through interference patterns and frequency shifts in the FSRS spectra.  For instance, an analysis of the early time FSRS line shapes and the spectral kinetics of the C$_{10}$ hydrogen-out-of-plane (HOOP) mode during the ultrafast isomerisation of rhodopsin has recently been carried out by McCamant.\cite{McCamant2011}  That work correlates the early ($<1$\,ps) line shapes with frequency changes during the Raman vibrational free induction decay showing the HOOP mode relaxing with a 140-fs decay time.  The heterodyne detection method employed by FSRS that allows interference of the optical fields during the free induction decay cause negative signals, leading to the `derivative' type spectral features observed in the time-resolved FSRS spectra of rhodopsin up to 500-fs delays.  Our data %, despite careful background subtraction, 
only portray such derivative features at the earliest time delay, indicating that free-induction decay following the electronic excitation and charge transfer is extremely rapid.  Our conclusion that rapid loss of coherence occurs in PCDTBT:PCBM is supported by recent work by Gao \& Grey that reports, for blends of poly(3-hexylthiophene) (P3HT) with PCBM, using resonance-Raman spectroscopy excitation profiles, that vibrational coherence within the C=C modes takes place on $\le 100$-fs timescales.\cite{Gao2013} The fact that we see clear signals of the cation forming on these timescales confirms the charge transfer reaction is occurring well within the few-hundred-femtosecond timescale. 
This structurally sensitive measurement thus establishes unambiguously that polarons in PCDTBT:PCBM blends are created about two orders of magnitude faster than exciton self-localisation in the neat polymer (see below).

The overall shape of the FSRS spectrum of PCDTBT:PCBM at $< 500$-fs timescales is very close to that of the relaxed structure and decays with the population dynamics (see Fig~\ref{fig:fig3}a and Supporting Information Fig.~S1), with the exception of the carbazole mode at 1513\,cm$^{-1}$ whose dynamics will be discussed later in this article. Since the spectral signature of polarons emerges in the relaxed structure form at timescales earlier than what we would expect from structural reorganisation, this hints that the conformation of the polymer emerging from the Franck-Condon region is already close to the relaxed hole conformation in the polymer. The structural similarity of the polaronic polymer backbone at early time and equilibrium implies a very small structural reorganisation, and we  propose that this facilitates the ultrafast charge transfer at the heterojunction. 
% as suggested by Bakulin et al,\cite{Bakulin2012} and in dye-sensetised TiO$_2$, by Frontiera~\textit{et~al}.\cite{Frontiera:fk} 

Having established that polarons appear on ultrafast timescales in PCDTBT:PCBM blends, we address whether they are readily observable in the neat polymer as well. The FSRS spectra of neat PCDTBT, reported in Fig.~\ref{fig:fig2}c, do not display polaronic signatures, excluding that species as a direct photoexcitation on the neat polymer on timescales shorter than 100\,ps. In particular, the 1364\,cm$^{-1}$ band is absent, and the relative intensity of the 1260\,cm$^{-1}$ band with respect to the 1429\,cm$^{-1}$ band is at least four times lower at 0.5\,ps. As the excitation decays, we do not observe the rise of the characteristic 1364\,cm$^{-1}$ band, indicating that there is no conversion of the initial excitation into polarons at later times. Therefore, polarons are not a dominant primary photoexcitation in neat PCDTBT. We thus assign the observed FSRS spectra in the neat PCDTBT film to intrachain singlet exciton as the primary photoexcitation. Indeed, Etzold~\textit{et~al}.\ assigned the photoinduced transient absorption band in neat PCDTBT, centred at the energy we use for our Raman pump (1.38\,eV), to singlet excitons,\cite{Etzold2011} while triplets were argued to form on timescales much longer than those explored in the present experiment.\cite{Reish2012} In addition, the calculated exciton Raman spectral line shape in PCDTBT from our TDDFT calculations is in remarkable agreement with the experimental data (see Fig.~S2 in Supporting Information), which further supports our assignments. 

These direct structural measurements have important implications on our understanding of `push-pull' co-polymers, which comprise electron-rich and electron-poor moieties within the polymer repeat unit. Indeed, an hypothesis to rationalise the generally high solar power-conversion efficiency of this class of co-polymers would be that they readily form loosely bound polaron-pairs upon photo\-excitation, without the aid of a fullerene acceptor.\cite{Tautz2012,Wiebeler2013} The evidence presented here in favour of the singlet exciton contradict this hypothesis. We underline that according to our calculations and other work,\cite{Reish2012, Banerji2012} the singlet exciton exhibits some degree of charge-transfer-like character. However, we insist that this picture is distinct from the loosely bound polaron pairs hypothetised for similar `push-pull' co-polymers as noted above. 

%%%%%%%%%%%%%%%%%%%%%%%%%%%%%%%%%%%%%%%%%%%%%%%%%%%%%%%%%
\begin{figure}
\caption{\textbf{$| \;$ Comparison of femtosecond stimulated Raman and transient absorption dynamics.}  Shown are the dynamics of the stimulated Raman peak areas (extracted by Gaussian curve fit) and the excited species population in (\textbf{a}) PCDTBT:PCBM and (\textbf{c}) PCDTBT films. The analyzed FSRS peaks for the neat material and the blend, along with the symbol key for the time-resolved data shown in panels (a) and (c), are shown in panels (\textbf{b}) and (\textbf{d}), respectively. The data are normalised at 4\,ps for panel (a) and at 0.6\,ps in panel (c). The dotted lines in panels (a) and (c) are meant as a guide to the eye. Vector diagrams showing vibrational motion associated with the corresponding analysed peaks in a dimer of PCDTBT, determined by ab-initio calculations, are displayed for the cation (\textbf{e}) and the neutral molecule (\textbf{f}).  Animations of those vibrational modes can be viewed as supplementary information on the Nature Materials website.
\label{fig:fig3}}
\end{figure}
%%%%%%%%%%%%%%%%%%%%%%%%%%%%%%%%%%%%%%%%%%%%%%%%%%%%%%%%%

The evolution of the FSRS spectral line shape in neat PCDTBT is complex, but is consistent with exciton self-localisation on a picosecond timescale. To appreciate the evolution of individual modes, we extract their kinetics by fitting FSRS peaks with a Gaussian curve and report their area as a function of time on Fig.~\ref{fig:fig3}c, comparing them with population dynamics extracted from transient absorption measurements probed at 1.2\,eV (1035\,nm). The most intense FSRS peak at 1429\,cm$^{-1}$, comprising mainly thiophene C=C stretching and carbazole C---N---C bending/C---C stretching, follows the exciton population dynamics. However, other bands deviate significantly from population dynamics. Specifically, the 1507-cm$^{-1}$ peak decays to 75\% of its initial intensity within 3\,ps, while the 1260\,cm$^{-1}$ band grows on a similar timescale. The 1507-cm$^{-1}$ band is a carbazole mode not visible in the ground state of neutral PCDTBT. However, our TDDFT calculations reveal that coupling with neighbouring thiophenes activates this mode. 
The 1260-cm$^{-1}$ band is characterised by C---H in-plane bending and C=C stretching in the benzene ring along with C---N stretching in the thiadiazole ring and hence is sensitive to changes in electronic density on this unit. 
A slow increase in intensity is also observed in other modes, such as the 1045\,cm$^{-1}$ thiophene C---H bending mode. 
The evidence provided by FSRS thus helps draw a picture of the exciton evolution following photoexcitation to the Franck-Condon region, where the wavefunction is initially delocalized along the polymer chain.\cite{Banerji:2013ej}  The rapid decay of the 1507-cm$^{-1}$ mode indicates that initial coupling of carbazole with thiophenes is lost upon shifting of electron density from carbazole to the benzothiadazole acceptor, evidenced by the growing intensity of the 1260\,cm$^{-1}$ band.
Localisation of the electronic density on the benzothiadiazole and thiophenes is consistent with the calculated localisation of the electronic density in the LUMO (see Fig. S13 in the Supporting Information). Therefore, exciton localisation (self-trapping) takes place on a 3\,ps timescale in the case of PCDTBT, which is in general agreement with the 1--10-ps timescale reported for exciton self-localisation in polymers.\cite{Kobayashi2000, De2008}
                
%    \subsubsection*{Small geometric reorganisation might promote ultrafast charge separation}
We now return to a discussion of the FSRS dynamics displayed in Figs.~\ref{fig:fig2}a and \ref{fig:fig2}b, in which we had pointed out that the overall shape of the FSRS spectrum of PCDTBT:PCBM at  early times is comparable to that of the relaxed structure (100\,ps and beyond) and decays with the population dynamics (see Fig.~\ref{fig:fig3}a and Supporting Information Fig.~S1 and S4). 
           %  1513 mode decay
As noted above, the only vibrational decay dynamics that we observe involve the decay of the mode at 1513\,cm$^{-1}$. In fact, the 1200\,cm$^{-1}$ mode displays similar behaviour, though its modest intensity makes quantitative analysis challenging. Interestingly, what these two modes have in common is that their band intensity depends on the extent of the coupling  between carbazole and its neighbouring thiophene. 

We can rationalise the decay dynamics of these modes by  a dynamic decoupling of carbazole and thiophene, possibly arising from evolution of the dihedral angle between these moieties. We also note that although the 1513-cm$^{-1}$ mode decays rapidly,
this decay is slower in the blend than for the 1507\,cm$^{-1}$ mode in the neat film (Fig.~\ref{fig:fig2}c). This suggests that the mechanism in the blend is distinct from the excitonic behaviour in the neat film. 
Insight on the significance of those dynamics can be gained by considering previous work by Pensack~\textit{et~al}.\ focusing on a carboxyl spectator mode in PCBM using  ultrafast solvatochromism assisted vibrational spectroscopy.\cite{Pensack2009JACS} These authors proposed that vibrational energy originally in Franck-Condon active modes in a polymer:PCBM electron transfer reaction redistributes on a few-picosecond timescale into a distribution of vibrational modes including those involved in electron transport between PCBM molecules.\cite{Pensack2009JACS} We speculate that the rapid relaxation of the 1513\,cm$^{-1}$ mode could be related to mobile hole polarons as they move away from the interface. 

            %Further exciton relaxation on 100ps timescale
In contrast to the modest relaxation of the polymer backbone in the presence of nascent hole-polarons, the overall spectral shape of the FSRS spectrum of PCDTBT in the presence of excitons evolves continuously during the first 100\,ps after photoexcitation and probably beyond, which indicates that after the initial fast exciton self-trapping, a slower structural relaxation takes place. This indication would explain fluorescence up-conversion measurements on PCDTBT thin films, which show that the photoluminescence spectrum red shifts to its relaxed state within 200\,ps.\cite{Banerji2010} In particular, we observe red-shifting of the 1260 and 1471-cm$^{-1}$ bands that correspond to benzothiadazole and carbazole motion likely due to change in the bond order of the backbone in the thiophene-benzothiadiazole-thiophene unit, and shift of the electron density from the carbazole to the acceptor moiety, lengthening thus the carbazole C---C bonds (see Fig. S13 in Supporting Information).  %Blue-shifting of the 1050 and 1430\,cm$^{-1}$ modes  on the other hand can be associated with increased planarization between the carbazole and thiophene units.
On the other hand, the 1429-cm$^{-1}$ thiophene and carbazole band blue-shifts significantly between 50 and 100\,ps, reflecting structural reorganisation on those timescales, involving possibly dihedral angle changes between the two units.
                
                %mini-conclusion for this part to drive our points home
%In addition to demonstrating good photovoltaic performance, PCDTBT:PCBM presents ultrafast charge separation with subsequent modest polymer backbone reorganisation. If these features are related, then a small geometric reorganisation might promote efficient ultrafast charge separation in polymeric solar cells.
        
%\section*{Discussion}
    The two key observations presented in this paper are that 
        \begin{inparaenum}[(\itshape i\upshape)]
        \item in PCDTBT:PCBM, ultrafast charge transfer happens prior to exciton localisation, as exemplified by the emergence of polaron resonance-Raman signatures on PCDTBT on timescales well within a few hundred femtoseconds; \label{BigRes1}
        \item structural reorganisation after polaron formation is negligable, as we do not observe significant spectral shifts or changes in linewidth during the first 50\,ps. \label{BigRes2}
        \end{inparaenum}
        
Result~(\textit{\ref{BigRes1}}) is strongly indicative of ballistic electron transfer following photoexcitation of PCDTBT in the blend, and that polarons, not charge-transfer excitons, are the photoproduct of such rapid dynamics. This implies that the corresponding electron on a PCBM phase, not detected by our FSRS measurements, must be separated over a sufficiently large distance on average in order for the polaron  on the polymer backbone to not evolve under the resulting Coulomb potential on subnanosecond timescales. In the dielectric environment defined by this class of materials, this separation must be in the order of several nanometers. In fact, G\'elinas \textit{et al.}\ have analysed signatures of a transient Stark effect in the photobleach component of a transient absorption measurement in molecular donor blends with PCBM, and have concluded that over $\sim100$\,fs the average electron-hole separation evolves to several nanometers,\cite{Gelinas:2013} in agreement with our conclusion. Such rapid, long-range charge transfer can be rationalised by exciton delocalisation phenomena,\cite{Mukamel:2000fk,Banerji:2013ej} which can generally persist over sufficiently long timescales to play a role in the charge separation dynamics of interest in this work. 
Understanding the mechanism of ultrafast charge separation is important, as it was demonstrated that even when electronic alignments are unfavourable to charge separation (e.g.\ in the presence of triplets), ultrafast kinetics can dominate and lead to very high internal quantum efficiencies.\cite{Schlenker2012}
Golden-rule formalisms applied to molecular electron transfer between localised asymptotic states cannot generally describe such rapid dynamics; however, when formulated in a delocalised basis set, such rapid dynamics can be readily described.\cite{Zhang:uq} Another possible mechanism under which ultrafast electron transfer can occur is bath-induced coupling of asymptotic eigenstates of the system.\cite{Bittner2013a,Bittner2013b} In this mechanism, the vertical, non-equilibrium exciton state transfers by resonant tunnelling to a delocalised charge-transfer state induced by a dynamic coupling due to common zero-point fluctuations within a vibrational noise spectrum of the system. Indeed, such spectacularly rapid charge separation points strongly to quantum coherence dynamics which are correlated to the dynamics of the molecular lattice,\cite{AndreaRozzi:2013ba} and has been argued to play a key role in energy transfer dynamics in biological systems.\cite{Scholes:2010ix} Either of these two mechanisms permit exciton separation over timescales that are much shorter than those in which a Marcus-theory description of ultrafast electron transfer is valid.
        
        %This paragraph is about the lack of reorganisation
Observation~(\textit{\ref{BigRes2}}) concerning the lack of structural reorganisation after polaron formation is surprising, since polaronic relaxation following ultrafast charge transfer is generally expected in these polymers composed of such complex repeat units.   
Since the polaron in PCDTBT:PCBM is formed within 200\,fs, the subsequent limited structural evolution over 50\,ps implies that the nascent hole is sufficiently far from the electron counterpart such that they are free from their mutual Coulomb potential. If the charge pair was Coulomb-bound, we would expect evolution of the polaron vibrational signatures as the two charges relax within their attractive potential. 
Similar to our findings, another FSRS study concerning a coumarin:TiO$_2$ system also observed the absence of significant reorganisation after charge formation.\cite{Frontiera2009} Rapid charge separation is expected in this organic:inorganic system, 
since electron injection occurs into delocalised states in a conduction band of the inorganic nanostructure. The remaining positive charge on the dye therefore does not  evolve under the influence of the electron potential. It is surprising that this seems to occur in PCDTBT:PCBM as well, given that such delocalised states in acceptor molecular aggregates are not expected \textit{a priori}. 
Regarding the modest structural evolution on the polymer, we note that \"{O}sterbacka~\textit{et~al}.\ concluded that an increase in polymer chain stiffness is related to a smaller polaron relaxation energy.\cite{Osterbacka2002} 
Similarly, Coffey~\textit{et~al}.\ have  suggested that a key route for improving future excitonic solar cells might be to reduce the reorganisation energy inherent to photoinduced electron transfer.\cite{Coffey2012} 
Indeed, we propose that correlating reorganisation dynamics with macroscopic optoelectronic parameters will enable a fundamental understanding of the critical factors that lead to efficient photovoltaic systems at a molecular level.
The modest polaron relaxation in PCDTBT:PCBM is an intriguing result, and one that opens a new door to understanding of the differences between material systems that produce efficient solar cells and those that should but do not perform as well, even if they obey the same `rational design' principles based on empirical design rules relating device parameters (e.g.\ open-circuit voltage, short-circuit photocurrent) to materials properties (e.g.\ absorption spectral range, reduction potentials with respect to PCBM, etc.). 

%this paragraph is about the 1-step VS 2-step mechanism.
Finally, we discuss that observations (\textit{\ref{BigRes1}}) and (\textit{\ref{BigRes2}}) indicate the possibility of parallel pathways for polaron formation in organic solar cells, as was also highlighted in the literature for P3HT:PCBM. Sheng~\textit{et~al}.\ demonstrated that polarons in P3HT:PCBM are created through two different channels: a direct, ultrafast one-step process and a slower, two-step process involving the creation and the dissociation of charge-transfer states.\cite{Sheng2012}  Gao and Grey inferred that the dephasing of the vibrational coherence on ultrafast timescales does not depend on PCBM loading ratio in regiorandom P3HT:PCBM,\cite{Gao2013} indicating that ultrafast coupling to charge-transfer states is not significant. Therefore, because photocarriers are generated with high yield in that system, there must be a direct mechanism for their generation that must bypass charge-transfer intermediates. 
Similarly, the present work points strongly to direct photogeneration (i.e.\ no CT intermediate) of unbound polarons in PCDTBT:PCBM, although it is also clear that CTX must form eventually in this system since in a previous publication we observed geminate recombination through CTX emission on nanosecond to microsecond timescales.\cite{Provencher:2012fm} We note from this work, however, that the direct polaron generation mechanism is operative on ultrafast timescales in PCDTBT:PCBM, with no forthright evidence of early involvement of intermediate charge-transfer states, while both channels may contribute to photocurrent in P3HT:PCBM. The degree of importance of the charge-transfer-state-mediated mechanism appears to be highly system dependent. 
%Therefore, %P3HT:PCBM and PCDTBT:PCBM both show multiple pathways for polaron generation, 
%the dominating channel is different in P3HT:PCBM and PCDTBT:PCBM systems. %But we just argued that the John Grey paper would be consistent with weak coupling to CT states. 
%In the former, the two-step process involving charge-transfer intermediate for polaron generation is desirable as these are more mobile that polarons generated directly,\cite{Sheng2012} but this pathway risks recombination to produce CTX, thus potentially hampering the efficiency. For the latter, the direct, one-step generation is the dominating pathway. 
In ultrafast pump-push photocurrent measurements, Bakulin~\textit{et~al}.\ demonstrated that once charge pairs relax into bound states, mid-infrared excitation promotes these to a near-conduction-edge state such that photocurrent is recovered in many polymer:PCBM and polymer:polymer blends, but intriguingly, not in PCDTBT:PC$_{70}$BM.\cite{Bakulin2012} 
If direct photocarrier generation is dominant in PCDTBT:PCBM and related efficient polymer systems,\cite{Grancini2012} avoiding the two-step process as is found in P3HT:PCBM\cite{Sheng2012} might be key in driving up efficiencies in solar cells by avoiding highly-bound CTX states,\cite{Gelinas:2011cp} even if it is the case that these states are photocurrent precursors.\cite{Vandewal:2013uq}
The conclusion that charge self-trapping is limited in PCDTBT blends is the key observation that must be understood in order to better engineer polymers for solar-cell applications.

\section*{Conclusion}
 We have implemented stimulated Raman spectroscopy on thin films of a photovoltaic-relevant polymer:fullerene heterostructure to reveal the structural dynamics of the polymer after photoexcitation, and assign the molecular origin of the observed features in transient absorption\cite{Tong2010} and time-resolved photoluminescence\cite{Banerji2010} spectroscopies. Upon absorption in the lowest ($\pi$, $\pi^{*}$) band at 2.21\,eV (560\,nm), ultrafast charge generation occurs within 200\,fs to form a hole polaron on the polymer. The charge separation takes place before exciton localisation (within 3\,ps) in the neat film, while the exciton is far from vibrational equilibrium. Furthermore, vibrational relaxation of the resulting polaron on the polymer is modest over the first 100\,ps, which suggests that the hole is free from the Coulomb potential that the twin negative charge would impart. This mechanistic insight is key for future optimization of the efficiency of polymer:fullerene based solar cells as it provides a window into the relationship between electronic dynamics and those of the polymer lattice involved in the conversion of excitons to photocarriers, which is of key importance for fundamental mechanistic understanding with molecular detail. Moreover, these conclusions are generally important in molecular photoexcitation because virtually every photochemical, photophysical, or spectroscopic process in molecular materials involves coupled dynamics of electrons and the structure.

\begin{methods}

\subsection{Sample preparation.}

PCDTBT ($M_n = 13,000$\,kg/mol; $M_w = 30,000$\,kg/mol; PDI = 2.3) was synthesised by Serge Beaupr\'e in the group of Mario Leclerc at Universit\'e Laval as described elsewhere\cite{Blouin:2007iy} and used as received. PCBM was purchased from Solenne (PCBM$_{60}$, 99.5\%, lot 15-05-12) and used as received. Thin films were wire-bar-coated at 100$^\circ$C on sapphire substrates (UQG Optics, 15-mm diameter, 1-mm thickness) using a bar with a channel width of 0.05\,mm from solutions in 1,2-dichlorobenzene (HPLC grade, Alfa Aesar) of neat PCDTBT at a concentration of 8\,mg/mL and PCDTBT:PCBM in a ratio of 1:4 at a concentration of 8\,mg of polymer for 32\,mg of PCBM in 1\,mL. The polymer concentration was kept constant constant to preserve the required viscosity of the solutions to make the films. Morphologicial studies including X-ray, optical and thermal experiments on this material system can be found in the Supporting Information, see Fig.~S5 and S6.

\subsection{Spectroscopic methods.}

Steady-state spontaneous Raman spectra were acquired with a Raman microscope (Renishaw inVia) with either 830 or 514-nm excitation. 
Femtosecond stimulated Raman spectroscopy (FSRS) was performed at the Central Laser Facility of the Rutherford-Appleton Laboratory using the ULTRA setup, a 10-kHz synchronized dual-arm femtosecond and picosecond laser system described elsewhere.\cite{Greetham2010} 
FSRS requires the generation of three pulses: the actinic pulse (560\,nm, $\sim 50$\,fs), the Raman pulse (900\,nm, 1.5\,ps) and the broadband probe pulse ($\sim 50$\,fs), which yields an experimental time resolution of $\sim 70$\,fs. These were generated using two custom titanium-sapphire amplifiers (Thales Laser) seeded from a single oscillator (20\,fs, Femtolasers) to create synchronised femtosecond and picosecond outputs after compression, respectively. 
The femtosecond beam was split into two: one femtosecond beam pumped an optical parametric amplifier (Light Conversion TOPAS) to generate the actinic pulse, while the other femtosecond beam was focused onto a lead-doped-glass window to create a broadband white light continuum to generate the broad-band probe pulse. The 800-nm fundamental wavelength of the laser was then rejected from the white light continuum by a pair of notch filters (Kaiser Holographic notch plus) to avoid two-photon absorption in the sample. These notch filters are responsible for the gap at 800\,nm in the transient absorption spectra presented in Fig.~\ref{fig:fig1}.The picosecond beam was tuned to 900\,nm by an optical parametric amplifier (Light Conversion TOPAS) and acted as the Raman pump pulse. The actinic and Raman beams were mechanically chopped at 5\,kHz and 2.5\,kHz, respectively, while the probe beam was kept at the system repetition rate of 10\,kHz. These different repetition rates create four different sequences of pulses, namely actinic-probe (01), Raman-probe (10), actinic-Raman-probe (11) and probe alone (00). The actinic-probe and probe alone sequences are used to obtain the transient absorption spectrum, while the FSRS spectra are obtained using a combination of those four sequences, as explained by Greetham \textit{et al.}\cite{Greetham2010} Temporal dispersion of the pulses was minimized using reflective optics to transport beams to the sample. A linear motor drive translation stage (Newport) provided femtosecond to nanosecond pump-probe timing.The beams were focused onto the sample to typical beam diameters of 50\,$\mu$m (probe beam) and 100\,$\mu$m (actinic and Raman beams) set in a non-collinear geometry with with an angle of 10$^{\circ}$ between the Raman and actinic beams and the probe.

Samples were placed in a cold-finger cryostat (Janis Research Company) and kept under a dynamic vacuum ($10^{-6}$\,mbar). Low temperatures were achieved by flowing liquid helium, reaching 100\,K and 180\,K for the PCDTBT:PCBM and the neat PCDTBT films, respectively. The cryostat was rastered by a mecanical stage in the plane perpendicular to the beams to minimize cumulative photo-damage. 

After passing through the sample, the probe beam was collimated and focused into a spectrograph (0.25\,m f/4 DK240, Spectral Products), then detected shot-by-shot using custom high-rate-readout linear dual detectors (Quantum Detectors). The spectra were averaged using a computer, averaging 216,000 detected shots per spectra for the neat PCDTBT film and 441,000 shots per spectra for the PCDTBT:\-PCBM film.

The spectra obtained were then smoothed using a feature-preserving Savitzky-Golay filter in MATLAB. The step size of the filter was small to avoid losing spectral accuracy. The FSRS spectra were baseline corrected, using a polynomial fit to the broad background shape, preserving the FSRS peaks (see Fig.~S7 in Supporting Information). FSRS peak dynamics (peak area, centre frequency) were extracted by fitting a single Gaussian curve to individual peaks, or multiple Gaussian curves in areas of peak congestion.

\subsection{Computational methods.}
All calculations were carried out within the framework of density functional theory (DFT)\cite{Martin2004} using Gaussian 03 and Gaussian 09 software\cite{Gaussian09} with the 6-311g(d) basis set.\cite{Krishnan1980} The functional B3LYP\cite{Becke1993} was used, which contains an empirically fitted percentage of exact-exchange for a better description of organic molecules. Other functionals were tested, such as CAM-B3LYP,\cite{Yanai2004} but the B3LYP was chosen due to a better description of the experimental vibrational spectrum of the ground state. All the optimization calculations were forced in a planar geometry, since the low-frequenecy torsion motion of the excited state threw the system out of the harmonic regime. Therefore, all torsion effects are not described by those calculations.

While the ground-state and cation properties are calculated with DFT, the optically-excited-state properties are obtained by applying time-dependent density functional theory (TDDFT).\cite{Marques2006} The new vibrational frequencies of each mode are obtained by fitting a parabola through nine single-point energy calculations around the energy minimum of the first singlet excited state, going from the minimum to the zero-point motion amplitude of the mode. This technique assumes that the effect on the frequency eigenvalues is limited to the first order, and that second-order effects of displacement eigenvectors can be neglected.

The excited state intensities depend on the charge reorganisation of the addition of an electron and a hole in our system, and they are approximated using the cation and anion ground state calculations. The ground-state mode was decomposed in the cation and anion vibrational modes, and a weighted average of all the intensities were done accordingly, with equal weight between the cation and the anion, to obtain the new intensity. This technique neglects the effect of the interaction between the electron and the hole in the excited state. This effect is presumed to be small in PCDTBT's case, since the description of the first singlet excited state after geometrical optimisation is composed of 97\% of the HOMO$\to$LUMO transition. However, the new excited vibrational intensities must only be used to indicate the tendencies of the effects of the charge reorganisation, and their quantitative significance is ambiguous.

\end{methods}

\newpage

%\bibliography{FSRS_bibliography}

\begin{thebibliography}{10}
\expandafter\ifx\csname url\endcsname\relax
  \def\url#1{\texttt{#1}}\fi
\expandafter\ifx\csname urlprefix\endcsname\relax\def\urlprefix{URL }\fi
\providecommand{\bibinfo}[2]{#2}
\providecommand{\eprint}[2][]{\url{#2}}

\bibitem{Clarke:2010gb}
\bibinfo{author}{Clarke, T.~M.} \& \bibinfo{author}{Durrant, J.~R.}
\newblock \bibinfo{title}{{Charge Photogeneration in Organic Solar Cells}}.
\newblock \emph{\bibinfo{journal}{Chem. Rev.}} \textbf{\bibinfo{volume}{110}},
  \bibinfo{pages}{6736--6767} (\bibinfo{year}{2010}).

\bibitem{Provencher:fk}
\bibinfo{author}{Provencher, F.}, \bibinfo{author}{Beljonne, D.},
  \bibinfo{author}{Br\'{e}das, J.-L.} \& \bibinfo{author}{Silva, C.}
\newblock \bibinfo{title}{Charge photogeneration dynamics in polymer solar
  cells}.
\newblock \emph{\bibinfo{journal}{Nat. Mater.}}
  \textbf{\bibinfo{volume}{submitted}} (\bibinfo{year}{2013}).

\bibitem{MorteaniPRL2004}
\bibinfo{author}{Morteani, A.~C.}, \bibinfo{author}{Sreearunothai, P.},
  \bibinfo{author}{Herz, L.~M.}, \bibinfo{author}{Friend, R.~H.} \&
  \bibinfo{author}{Silva, C.}
\newblock \bibinfo{title}{{Exciton Regeneration at Polymeric Semiconductor
  Heterojunctions}}.
\newblock \emph{\bibinfo{journal}{Phys. Rev. Lett.}}
  \textbf{\bibinfo{volume}{92}}, \bibinfo{pages}{247402}
  (\bibinfo{year}{2004}).

\bibitem{Hallermann:2008dj}
\bibinfo{author}{Hallermann, M.}, \bibinfo{author}{Haneder, S.} \&
  \bibinfo{author}{Da~Como, E.}
\newblock \bibinfo{title}{{Charge-transfer states in conjugated
  polymer/fullerene blends: Below-gap weakly bound excitons for polymer
  photovoltaics}}.
\newblock \emph{\bibinfo{journal}{Appl. Phys. Lett.}}
  \textbf{\bibinfo{volume}{93}}, \bibinfo{pages}{053307}
  (\bibinfo{year}{2008}).

\bibitem{Bakulin2009}
\bibinfo{author}{Bakulin, A.~A.}, \bibinfo{author}{Martyanov, D.},
  \bibinfo{author}{Paraschuk, D.~Y.}, \bibinfo{author}{van Loosdrecht, P.
  H.~M.} \& \bibinfo{author}{Pshenichnikov, M.~S.}
\newblock \bibinfo{title}{{Charge-transfer complexes of conjugated polymers as
  intermediates in charge photogeneration for organic photovoltaics}}.
\newblock \emph{\bibinfo{journal}{Chem. Phys. Lett.}}
  \textbf{\bibinfo{volume}{482}}, \bibinfo{pages}{99--104}
  (\bibinfo{year}{2009}).

\bibitem{Tong2010}
\bibinfo{author}{Tong, M.} \emph{et~al.}
\newblock \bibinfo{title}{{Charge carrier photogeneration and decay dynamics in
  the poly(2,7-carbazole) copolymer PCDTBT and in bulk heterojunction
  composites with PC$_{70}$BM}}.
\newblock \emph{\bibinfo{journal}{Phys. Rev. B}} \textbf{\bibinfo{volume}{81}},
  \bibinfo{pages}{125210} (\bibinfo{year}{2010}).

\bibitem{Grancini2012}
\bibinfo{author}{Grancini, G.} \emph{et~al.}
\newblock \bibinfo{title}{{Hot exciton dissociation in polymer solar cells}}.
\newblock \emph{\bibinfo{journal}{Nat. Mater.}} \textbf{\bibinfo{volume}{12}},
  \bibinfo{pages}{29--33} (\bibinfo{year}{2012}).

\bibitem{Jailaubekov2012}
\bibinfo{author}{Jailaubekov, A.~E.} \emph{et~al.}
\newblock \bibinfo{title}{{Hot charge-transfer excitons set the time limit for
  charge separation at donor/acceptor interfaces in organic photovoltaics}}.
\newblock \emph{\bibinfo{journal}{Nat. Mater.}} \textbf{\bibinfo{volume}{11}},
  \bibinfo{pages}{1--8} (\bibinfo{year}{2012}).

\bibitem{Hallermann:2009cq}
\bibinfo{author}{Hallermann, M.} \emph{et~al.}
\newblock \bibinfo{title}{{Charge Transfer Excitons in Polymer/Fullerene
  Blends: The Role of Morphology and Polymer Chain Conformation}}.
\newblock \emph{\bibinfo{journal}{Adv. Funct. Mater.}}
  \textbf{\bibinfo{volume}{19}}, \bibinfo{pages}{3662--3668}
  (\bibinfo{year}{2009}).

\bibitem{Gelinas:2011cp}
\bibinfo{author}{G\'elinas, S.} \emph{et~al.}
\newblock \bibinfo{title}{{The Binding Energy of Charge-Transfer Excitons
  Localized at Polymeric Semiconductor Heterojunctions}}.
\newblock \emph{\bibinfo{journal}{J. Phys. Chem. C}}
  \textbf{\bibinfo{volume}{115}}, \bibinfo{pages}{7114--7119}
  (\bibinfo{year}{2011}).

\bibitem{Bakulin2012}
\bibinfo{author}{Bakulin, A.~A.} \emph{et~al.}
\newblock \bibinfo{title}{{The Role of Driving Energy and Delocalized States
  for Charge Separation in Organic Semiconductors}}.
\newblock \emph{\bibinfo{journal}{Science}} \textbf{\bibinfo{volume}{335}},
  \bibinfo{pages}{1340--1344} (\bibinfo{year}{2012}).

\bibitem{Lee2010}
\bibinfo{author}{Lee, J.} \emph{et~al.}
\newblock \bibinfo{title}{{Charge Transfer State Versus Hot Exciton
  Dissociation in Polymer− Fullerene Blended Solar Cells}}.
\newblock \emph{\bibinfo{journal}{J. Am. Chem. Soc.}}
  \textbf{\bibinfo{volume}{132}}, \bibinfo{pages}{11878--11880}
  (\bibinfo{year}{2010}).

\bibitem{Vandewal:2013uq}
\bibinfo{author}{Vandewal, K.} \emph{et~al.}
\newblock \bibinfo{title}{Efficient charge generation by relaxed
  charge-transfer states at organic interfaces}.
\newblock \emph{\bibinfo{journal}{Nat. Mater.}}
  \textbf{\bibinfo{volume}{Submitted}} (\bibinfo{year}{2013}).

\bibitem{Pensack2009JACS}
\bibinfo{author}{Pensack, R.~D.} \& \bibinfo{author}{Asbury, J.~B.}
\newblock \bibinfo{title}{{Barrierless Free Carrier Formation in an Organic
  Photovoltaic Material Measured with Ultrafast Vibrational Spectroscopy}}.
\newblock \emph{\bibinfo{journal}{J. Am. Chem. Soc.}}
  \textbf{\bibinfo{volume}{131}}, \bibinfo{pages}{15986--15987}
  (\bibinfo{year}{2009}).

\bibitem{McCamant2004}
\bibinfo{author}{McCamant, D.~W.}, \bibinfo{author}{Kukura, P.},
  \bibinfo{author}{Yoon, S.} \& \bibinfo{author}{Mathies, R.~A.}
\newblock \bibinfo{title}{{Femtosecond broadband stimulated Raman spectroscopy:
  Apparatus and methods}}.
\newblock \emph{\bibinfo{journal}{Rev. Sci. Instrum.}}
  \textbf{\bibinfo{volume}{75}}, \bibinfo{pages}{4971} (\bibinfo{year}{2004}).

\bibitem{Park2009}
\bibinfo{author}{Park, S.} \emph{et~al.}
\newblock \bibinfo{title}{{Bulk heterojunction solar cells with internal
  quantum efficiency approaching 100\%}}.
\newblock \emph{\bibinfo{journal}{Nat. Photon.}} \textbf{\bibinfo{volume}{3}},
  \bibinfo{pages}{297--302} (\bibinfo{year}{2009}).

\bibitem{Sun2011}
\bibinfo{author}{Sun, Y.}, \bibinfo{author}{Takacs, C.~J.},
  \bibinfo{author}{Cowan, S.~R.}, \bibinfo{author}{Seo, J.~H.} \&
  \bibinfo{author}{Gong, X.}
\newblock \bibinfo{title}{{Efficient, Air-Stable Bulk Heterojunction Polymer
  Solar Cells}}.
\newblock \emph{\bibinfo{journal}{Adv. Mater.}} \textbf{\bibinfo{volume}{23}},
  \bibinfo{pages}{2226--2230} (\bibinfo{year}{2011}).

\bibitem{Etzold2011}
\bibinfo{author}{Etzold, F.} \emph{et~al.}
\newblock \bibinfo{title}{{Ultrafast Exciton Dissociation Followed by
  Nongeminate Charge Recombination in PCDTBT:PCBM Photovoltaic Blends}}.
\newblock \emph{\bibinfo{journal}{J. Am. Chem. Soc.}}
  \textbf{\bibinfo{volume}{133}}, \bibinfo{pages}{9469--9479}
  (\bibinfo{year}{2011}).

\bibitem{Aich2009}
\bibinfo{author}{Badrou~A\"{i}ch, R.}, \bibinfo{author}{Blouin, N.},
  \bibinfo{author}{Bouchard, A.} \& \bibinfo{author}{Leclerc, M.}
\newblock \bibinfo{title}{{Electrical and Thermoelectric Properties of
  Poly(2,7-Carbazole) Derivatives}}.
\newblock \emph{\bibinfo{journal}{Chem. Mater.}} \textbf{\bibinfo{volume}{21}},
  \bibinfo{pages}{751--757} (\bibinfo{year}{2009}).

\bibitem{Wiebeler2013}
\bibinfo{author}{Wiebeler, C.} \emph{et~al.}
\newblock \bibinfo{title}{{Spectral Signatures of Polarons in Conjugated
  Co-polymers}}.
\newblock \emph{\bibinfo{journal}{J. Phys. Chem. B}}
  \textbf{\bibinfo{volume}{117}}, \bibinfo{pages}{4454--4460}
  (\bibinfo{year}{2013}).

\bibitem{Reish2012}
\bibinfo{author}{Reish, M.~E.}, \bibinfo{author}{Nam, S.},
  \bibinfo{author}{Lee, W.}, \bibinfo{author}{Woo, H.~Y.} \&
  \bibinfo{author}{Gordon, K.~C.}
\newblock \bibinfo{title}{{A Spectroscopic and DFT Study of the Electronic
  Properties of Carbazole-Based D--A Type Copolymers}}.
\newblock \emph{\bibinfo{journal}{J. Phys. Chem. C}}
  \textbf{\bibinfo{volume}{116}}, \bibinfo{pages}{21255--21266}
  (\bibinfo{year}{2012}).

\bibitem{McCamant2011}
\bibinfo{author}{McCamant, D.~W.}
\newblock \bibinfo{title}{{Re-Evaluation of Rhodopsin's Relaxation Kinetics
  Determined from Femtosecond Stimulated Raman Lineshapes}}.
\newblock \emph{\bibinfo{journal}{J. Phys. Chem. B}}
  \textbf{\bibinfo{volume}{115}}, \bibinfo{pages}{9299--9305}
  (\bibinfo{year}{2011}).

\bibitem{Gao2013}
\bibinfo{author}{Gao, J.} \& \bibinfo{author}{Grey, J.~K.}
\newblock \bibinfo{title}{{Resonance Raman overtones reveal vibrational
  displacements and dynamics of crystalline and amorphous
  poly(3-hexylthiophene) chains in fullerene blends}}.
\newblock \emph{\bibinfo{journal}{J. Chem. Phys.}}
  \textbf{\bibinfo{volume}{139}}, \bibinfo{pages}{044903}
  (\bibinfo{year}{2013}).

\bibitem{Tautz2012}
\bibinfo{author}{Tautz, R.} \emph{et~al.}
\newblock \bibinfo{title}{{Structural correlations in the generation of polaron
  pairs in low-bandgap polymers for photovoltaics}}.
\newblock \emph{\bibinfo{journal}{Nat. Commun.}} \textbf{\bibinfo{volume}{3}},
  \bibinfo{pages}{970--8} (\bibinfo{year}{2012}).

\bibitem{Banerji2012}
\bibinfo{author}{Banerji, N.} \emph{et~al.}
\newblock \bibinfo{title}{{Breaking Down the Problem: Optical Transitions,
  Electronic Structure, and Photoconductivity in Conjugated Polymer PCDTBT and
  in Its Separate Building Blocks}}.
\newblock \emph{\bibinfo{journal}{J. Phys. Chem. C}}
  \textbf{\bibinfo{volume}{116}}, \bibinfo{pages}{11456--11469}
  (\bibinfo{year}{2012}).

\bibitem{Banerji:2013ej}
\bibinfo{author}{Banerji, N.}
\newblock \bibinfo{title}{{Sub-picosecond delocalization in the excited state
  of conjugated homopolymers and donor--acceptor copolymers}}.
\newblock \emph{\bibinfo{journal}{J. Mater. Chem. C}}
  \textbf{\bibinfo{volume}{1}}, \bibinfo{pages}{3052} (\bibinfo{year}{2013}).

\bibitem{Kobayashi2000}
\bibinfo{author}{Kobayashi, T.} \emph{et~al.}
\newblock \bibinfo{title}{{Self-trapped exciton dynamics in highly ordered and
  disordered films of polythiophene derivative}}.
\newblock \emph{\bibinfo{journal}{Phys. Rev. B}} \textbf{\bibinfo{volume}{62}},
  \bibinfo{pages}{8580} (\bibinfo{year}{2000}).

\bibitem{De2008}
\bibinfo{author}{De, S.} \emph{et~al.}
\newblock \bibinfo{title}{{Exciton dynamics in alternating
  polyfluorene/fullerene blends}}.
\newblock \emph{\bibinfo{journal}{Chem. Phys.}} \textbf{\bibinfo{volume}{350}},
  \bibinfo{pages}{14--22} (\bibinfo{year}{2008}).

\bibitem{Banerji2010}
\bibinfo{author}{Banerji, N.}, \bibinfo{author}{Cowan, S.~R.},
  \bibinfo{author}{Leclerc, M.}, \bibinfo{author}{Vauthey, E.} \&
  \bibinfo{author}{Heeger, A.~J.}
\newblock \bibinfo{title}{{Exciton Formation, Relaxation, and Decay in
  PCDTBT}}.
\newblock \emph{\bibinfo{journal}{J. Am. Chem. Soc.}}
  \textbf{\bibinfo{volume}{132}}, \bibinfo{pages}{17459--17470}
  (\bibinfo{year}{2010}).

\bibitem{Gelinas:2013}
\bibinfo{author}{G\'elinas, S.} \emph{et~al.}
\newblock \bibinfo{title}{Ultrafast long-range charge separation in organic
  semiconductor photovoltaic diodes}.
\newblock \emph{\bibinfo{journal}{Science}}
  \textbf{\bibinfo{volume}{Submitted}} (\bibinfo{year}{2013}).

\bibitem{Mukamel:2000fk}
\bibinfo{author}{Mukamel, S.}
\newblock \bibinfo{title}{Multidimensional femtosecond correlation
  spectroscopies of electronic and vibrational excitations}.
\newblock \emph{\bibinfo{journal}{Annu. Rev. Phys. Chem.}}
  \textbf{\bibinfo{volume}{51}}, \bibinfo{pages}{691--729}
  (\bibinfo{year}{2000}).

\bibitem{Schlenker2012}
\bibinfo{author}{Schlenker, C.~W.} \emph{et~al.}
\newblock \bibinfo{title}{{Polymer Triplet Energy Levels Need Not Limit
  Photocurrent Collection in Organic Solar Cells}}.
\newblock \emph{\bibinfo{journal}{J. Am. Chem. Soc.}}
  \textbf{\bibinfo{volume}{134}}, \bibinfo{pages}{19661--19668}
  (\bibinfo{year}{2012}).

\bibitem{Zhang:uq}
\bibinfo{author}{Zhang, W.-M.}, \bibinfo{author}{Meier, T.},
  \bibinfo{author}{Chernyak, V.} \& \bibinfo{author}{Mukamel, S.}
\newblock \bibinfo{title}{Exciton-migration and three-pulse femtosecond optical
  spectroscopies of photosynthetic antenna complexes}.
\newblock \emph{\bibinfo{journal}{J. Chem. Phys.}}
  \textbf{\bibinfo{volume}{108}}, \bibinfo{pages}{108--120}
  (\bibinfo{year}{1998}).

\bibitem{Bittner2013a}
\bibinfo{author}{Bittner, E.~R.} \& \bibinfo{author}{Silva, C.}
\newblock \bibinfo{title}{Noise-induced quantum coherence drives photocarrier
  generation dynamics at polymeric semiconductor heterojunctions}.
\newblock \emph{\bibinfo{journal}{Nat. Commun.}}
  \textbf{\bibinfo{volume}{Submitted}} (\bibinfo{year}{2013}).

\bibitem{Bittner2013b}
\bibinfo{author}{Bittner, E.~R.} \& \bibinfo{author}{Silva, C.}
\newblock \bibinfo{title}{Exciton fission via long-range resonant tunnelling in
  organic photovoltaics}.
\newblock \emph{\bibinfo{journal}{Phys. Rev. X}}
  \textbf{\bibinfo{volume}{Submitted}} (\bibinfo{year}{2013}).

\bibitem{AndreaRozzi:2013ba}
\bibinfo{author}{Rozzi, A.} \emph{et~al.}
\newblock \bibinfo{title}{{Quantum coherence controls the charge separation in
  a prototypical artificial light-harvesting system}}.
\newblock \emph{\bibinfo{journal}{Nat. Commun.}} \textbf{\bibinfo{volume}{4}},
  \bibinfo{pages}{1602} (\bibinfo{year}{2013}).

\bibitem{Scholes:2010ix}
\bibinfo{author}{Scholes, G.~D.}
\newblock \bibinfo{title}{{Quantum-Coherent Electronic Energy Transfer: Did
  Nature Think of It First?}}
\newblock \emph{\bibinfo{journal}{J. Phys. Chem. Lett.}}
  \textbf{\bibinfo{volume}{1}}, \bibinfo{pages}{2--8} (\bibinfo{year}{2010}).

\bibitem{Frontiera2009}
\bibinfo{author}{Frontiera, R.~R.}, \bibinfo{author}{Dasgupta, J.} \&
  \bibinfo{author}{Mathies, R.~A.}
\newblock \bibinfo{title}{{Probing Interfacial Electron Transfer in Coumarin
  343 Sensitized TiO 2Nanoparticles with Femtosecond Stimulated Raman}}.
\newblock \emph{\bibinfo{journal}{J. Am. Chem. Soc.}}
  \textbf{\bibinfo{volume}{131}}, \bibinfo{pages}{15630--15632}
  (\bibinfo{year}{2009}).

\bibitem{Osterbacka2002}
\bibinfo{author}{{\"O}sterbacka, R.}, \bibinfo{author}{Jiang, X.},
  \bibinfo{author}{An, C.}, \bibinfo{author}{Horovitz, B.} \&
  \bibinfo{author}{Vardeny, Z.~V.}
\newblock \bibinfo{title}{{Photoinduced quantum interference antiresonances in
  {$\pi$}-conjugated polymers}}.
\newblock \emph{\bibinfo{journal}{Phys. Rev. Lett.}}
  \textbf{\bibinfo{volume}{88}}, \bibinfo{pages}{226401}
  (\bibinfo{year}{2002}).

\bibitem{Coffey2012}
\bibinfo{author}{Coffey, D.~C.} \emph{et~al.}
\newblock \bibinfo{title}{{An Optimal Driving Force for Converting Excitons
  into Free Carriers in Excitonic Solar Cells}}.
\newblock \emph{\bibinfo{journal}{J. Phys. Chem. C}}
  \textbf{\bibinfo{volume}{116}}, \bibinfo{pages}{8916--8923}
  (\bibinfo{year}{2012}).

\bibitem{Sheng2012}
\bibinfo{author}{Sheng, C.-X.}, \bibinfo{author}{Basel, T.},
  \bibinfo{author}{Pandit, B.} \& \bibinfo{author}{Vardeny, Z.~V.}
\newblock \bibinfo{title}{{Photoexcitation dynamics in polythiophene/fullerene
  blends for photovoltaic applications}}.
\newblock \emph{\bibinfo{journal}{Organic Electronics}}
  \textbf{\bibinfo{volume}{13}}, \bibinfo{pages}{1031--1037}
  (\bibinfo{year}{2012}).

\bibitem{Provencher:2012fm}
\bibinfo{author}{Provencher, F.} \emph{et~al.}
\newblock \bibinfo{title}{{Slow geminate-charge-pair recombination dynamics at
  polymer: Fullerene heterojunctions in efficient organic solar cells}}.
\newblock \emph{\bibinfo{journal}{Journal of Polymer Science: Part B: Polymer
  Physics}} \textbf{\bibinfo{volume}{50}}, \bibinfo{pages}{1395--1404}
  (\bibinfo{year}{2012}).

\bibitem{Blouin:2007iy}
\bibinfo{author}{Blouin, N.}, \bibinfo{author}{Michaud, A.} \&
  \bibinfo{author}{Leclerc, M.}
\newblock \bibinfo{title}{{A Low-Bandgap Poly(2,7-Carbazole) Derivative for Use
  in High-Performance Solar Cells}}.
\newblock \emph{\bibinfo{journal}{Adv. Mater.}} \textbf{\bibinfo{volume}{19}},
  \bibinfo{pages}{2295--2300} (\bibinfo{year}{2007}).

\bibitem{Greetham2010}
\bibinfo{author}{Greetham, G.~M.} \emph{et~al.}
\newblock \bibinfo{title}{{ULTRA: A Unique Instrument for Time-Resolved
  Spectroscopy}}.
\newblock \emph{\bibinfo{journal}{Appl. Spectroscopy}}
  \textbf{\bibinfo{volume}{64}}, \bibinfo{pages}{1311--1319}
  (\bibinfo{year}{2010}).

\bibitem{Martin2004}
\bibinfo{author}{Martin, R.~M.}
\newblock \emph{\bibinfo{title}{{\emph{Electronic Structure: Basic Theory and
  Practical Methods}}}} (\bibinfo{publisher}{Cambridge University Press},
  \bibinfo{address}{New York}, \bibinfo{year}{2004}).

\bibitem{Gaussian09}
\bibinfo{author}{Frisch, M.~J.} \emph{et~al.}
\newblock \emph{\bibinfo{title}{{G}aussian 09}}.
\newblock \bibinfo{organization}{Gaussian, Inc.},
  \bibinfo{address}{Wallingford, CT} (\bibinfo{year}{2009}).

\bibitem{Krishnan1980}
\bibinfo{author}{Krishnan, R.}, \bibinfo{author}{Binkley, J.~S.},
  \bibinfo{author}{Seeger, R.} \& \bibinfo{author}{Pople, J.~A.}
\newblock \bibinfo{title}{Self-consistent molecular orbital methods. {XX}. {A}
  basis set for correlated wave functions}.
\newblock \emph{\bibinfo{journal}{J. Chem. Phys.}}
  \textbf{\bibinfo{volume}{72}}, \bibinfo{pages}{650--654}
  (\bibinfo{year}{1980}).

\bibitem{Becke1993}
\bibinfo{author}{Becke, A.~D.}
\newblock \bibinfo{title}{Density-functional thermochemistry. {III}. {T}he role
  of exact exchange}.
\newblock \emph{\bibinfo{journal}{J. Chem. Phys.}}
  \textbf{\bibinfo{volume}{98}}, \bibinfo{pages}{5648--5652}
  (\bibinfo{year}{1993}).

\bibitem{Yanai2004}
\bibinfo{author}{Yanai, T.}, \bibinfo{author}{Tew, D.~P.} \&
  \bibinfo{author}{Handy, N.~C.}
\newblock \bibinfo{title}{A new hybrid exchange-correlation functional using
  the {C}oulomb-attenuating method ({CAM-B3LYP})}.
\newblock \emph{\bibinfo{journal}{Chem. Phys. Lett.}}
  \textbf{\bibinfo{volume}{393}}, \bibinfo{pages}{51 -- 57}
  (\bibinfo{year}{2004}).

\bibitem{Marques2006}
\bibinfo{author}{Marques, M. A.~L.}
\newblock \emph{\bibinfo{title}{{\emph{Time-Dependent Density Functional
  Theory}}}} (\bibinfo{publisher}{Springer}, \bibinfo{address}{Berlin},
  \bibinfo{year}{2006}).

\end{thebibliography}

%% Here is the endmatter stuff: Supplementary Info, etc.
%% Use \item's to separate, default label is "Acknowledgements"

\newpage 

\begin{addendum}
 \item CS acknowledges funding from the Natural Sciences and Engineering Research Council of Canada, the Canada Research Chair in Organic Semiconductor Materials, the Royal Society, and the Leverhulme Trust. CS, SCH and NS acknowledge funding from Laserlab 2 EC Grant Agreement No.\ 228334 and STFC ULTRA facility for experimental time.  NS and CH acknowledge funding from the UK‚ Engineering and Physical Sciences Research Council EP/G060738/1 grant, the European Research Council (ERC) Starting Independent Research Fellowship under the grant agreement No. 279587 and King Abdullah University of Science and Technology (KAUST CRG).  The authors acknowledge the assistance of A. Scaccabarozzi for film deposition and are grateful to Serge Beaupr\'e and Mario Leclerc for providing the PCDTBT. 
  \item[Authors contribution] Samples were prepared by CH and FP. Microstructural characterisation was carried out by CH. Spectroscopic data were collected by FP, SCH, AWP, GMG, and MT, and analysed by FP and SCH. The experiments were conceived by CS, SCH, FP and AWP. Ab-initio calculations were carried out by NB. CS and SCH supervised the spectroscopy activity, MC supervised the theoretical calculations, and NS supervised the processing and characterisation activity. FP, SCH, and CS were primarily responsible for writing the manuscript, but all authors contributed to it.
 \item[Competing financial interests] The authors declare that they have no competing financial interests.
 \item[Additional information] Supplementary information is available in the online version of the paper. Reprints and permissions information is available online at www.nature.com/reprints. Correspondence and requests for materials should be addressed to SCH~(email: shayes@ucy.ac.cy) and CS~(email: carlos.silva@umontreal.ca).
\end{addendum}

%%
%% TABLES
%%
%% If there are any tables, put them here.
%%

\begin{table}
\footnotesize
 \begin{tabular}{c c c c c l }
    \hline
Spontaneous &  0.5\,ps& 100\,ps & DFT & TDDFT &Qualitative assignment\\
 pristine&  pristine & pristine & ground state&excited state&\\
(cm$^{-1}$) &(cm$^{-1}$)&(cm$^{-1}$)&(cm$^{-1}$)&(cm$^{-1}$)& \\
 \hline
 1062& 1057 &1067&1096&1082&  L : Th  CH ip $\delta$ + $\nu_{C=C}$\\
1135&1124 &1122&1153&1152&L: Cz CH ip $\delta$ + ring def. \\
1205&---&---&1231&1238&L : BT +Th CH ip $\delta$, BT ring def.  \\
1270 & 1260 &1256* & 1325+1327&1319+1308&D: BT  $\nu_{C-N}$  + $\nu_{C=C}$+ CH ip $\delta$, \\
 & & & & &Cz +Th CH ip $\delta$ + $\nu_{C=C}$\\
1348&---&---&1384 &1377&D: BT+ Th CH  $\delta$, Cz+ BT $\nu_{C-C}$ \\
1370&---&---&1405&1416&D : BT + Th + Cz $\nu_{C=C}$ + CH ip  $\delta$, \\
 & & & & &Cz $\nu_{C-N}$\\
1444 &1423&1429&1471+1485&1470&L: Cz, Th $\nu_{C=C}$ + CH ip $\delta$, Cz $\nu_{C-N}$  \\
1490&1473 &1457 & 1522+1527 & 1521 & L: Cz sym.  $\nu_{C=C}$, CH ip $\delta$ \\
---&1508&---&1543&1540&L : Cz asym. $\nu_{C=C}$, CH ip $\delta$ \\
1525 &---& ---&1569&1563&L: Cz asym. $\nu_{C=C}$+ CH ip  $\delta$ + $\nu_{C-N}$ \\
1540&---&---&1570+1572&1560+1556&L: BT + Th sym. $\nu_{C=C}$,  CH ip  $\delta$ \\
1570&1571&1555&1605 &1590&L: Cz $\nu_{C=C}$ +  CH ip $\delta$ \\
1622& 1608&1612&1664&1638& L: Cz $\nu_{C=C}$  +  CH ip $\delta$\\
 \hline
 \end{tabular}
 \label{tab:parameters}
 \caption{Neat and pristine (undoped, as cast) PCDTBT film  spontaneous Raman and FSRS shifts. Mode abbreviations: ip\,=\,in plane,  $\nu_{a-b}$\,=\,stretch of $a$---$b$ bond, $\delta$\,=\,bend, sym.\,=\,symmetric, asym.\,=\,asymmetric, def.\,=\,deformation, BT: benzothiadiazole, Cz: carbazole, Th: thiophene, L: localised, D: delocalised. *The peak at 1256\,cm$^{-1}$ is asymmetric (See Fig.~S11 in Supporting Information for vector graphics of the vibrational modes.)}
 \end{table}

\begin{table}
\footnotesize
 \begin{tabular}{c c c c l }
    \hline
Spontaneous &  0.5\,ps& 100\,ps & DFT & Qualitative assignment\\
doped & blend& blend & cation & \\
(cm$^{-1}$) &(cm$^{-1}$)&(cm$^{-1}$)&(cm$^{-1}$)& \\
 \hline
 1064&1083  &1083&1104&  L: Th  CH ip $\delta$ + $\nu_{C=C}$ \\
1130& 1129 &1146&1153&L: Cz CH ip $\delta$ + $\nu_{C=C}$\\
1200&1192&1190*&1182&L: Cz CH ip $\delta$  \\
1268 &  1262&1260 & 1303&D: BT  $\nu_{C-N}$ + CH ip $\delta$  + $\nu_{C=C}$, Cz +Th  CH ip $\delta$ \\
1334&---&---&1344&L: Cz ring deformation, Cz + Th CH ip $\delta$\\
1349&1357&1354&1371&D: Cz + BT $\nu_{C=C}$ + CH ip $\delta$ \\
1371&---&---&1397&D: BT + Cz + Th $\nu_{C=C}$ \\
1442 &1429&1434&1464&L: Th $\nu_{C=C}$, Cz $\nu_{C-N}$ +  $\nu_{C=C}$ + CH ip $\delta$ \\
1522 &1513 &1513**&1540+1547& L: Cz+Th asym. ring deformation, + CH ip  $\delta$ \\
1538&---&---&1560&L: BT sym.   $\nu_{C-C}$  + CH ip  $\delta$ , Th. asym.  $\nu_{C=C}$, \\
 & & & & CH ip  $\delta$  \\
1620& ---&--- &1648&L: Cz sym. ring deformation + ip CH $\delta$\\
 \hline
 \end{tabular}
 \label{tab:parameters}
 \caption{Doped PCDTBT film spontaneous resonance Raman at 830 nm (resonant with a positive polaron optical transition) and PCDTBT:PCBM (as-cast) film transient Raman shifts.  Mode abbreviations: ip\,=\,in plane, $\nu_{a-b}$\,=\,stretch of $a$---$b$ bond, $\delta$\,=\,bend,  sym.\,=\,symmetric, asym\,=\,asymmetric, BT: benzothiadiazole, Cz: carbazole, Th: thiophene, L: localised,D: delocalised.  *1190\,cm$^{-1}$ at 20\,ps,  **1513\,cm$^{-1}$ at 50\,ps. (See Fig.~S12 in Supporting Information for vector graphics of the vibrational modes.)}
 \end{table}

\end{document}